\documentclass[10pt,twocolumn,english,aps,prl,floatfix]{revtex4-1}
\usepackage{lmodern}

\usepackage[T1]{fontenc}
\usepackage[latin9]{inputenc}
\usepackage{babel}
\usepackage{float}
\usepackage{amsmath}
\usepackage{graphicx}
\usepackage[unicode=true,
 bookmarks=false,
 breaklinks=false,pdfborder={0 0 1},colorlinks=false,colorlinks=true,
  urlcolor=blue,
  linkcolor=blue,
  citecolor=blue]{hyperref}

\makeatletter
\usepackage{lmodern}
\usepackage[T1]{fontenc}
\usepackage{prettyref}
\usepackage{textgreek}
\usepackage{inputenc}
\usepackage{slashed}
\usepackage{hyperref}
\usepackage{bibentry}
\usepackage{soul}
\usepackage{color}
\usepackage{bm}

\makeatother

\begin{document}
\title{Quantum radiation reaction in aligned crystals beyond the local constant field approximation}
\author{T. N. Wistisen\textsuperscript{2}$ ^,$\textsuperscript{1}, A. Di Piazza\textsuperscript{2}, C. F. Nielsen\textsuperscript{1}, A. H. S{\o}rensen\textsuperscript{1}, and U. I. Uggerh{\o}j\textsuperscript{1}}
\affiliation{\textsuperscript{1}Department of Physics and Astronomy, Aarhus University,
8000 Aarhus, Denmark}
\affiliation{\textsuperscript{2}Max-Planck-Institut f{\"u}r Kernphysik, Saupfercheckweg
1, D-69117, Germany}
\begin{abstract}
We report on experimental spectra of photons radiated by 50 GeV positrons crossing silicon single crystals of thicknesses 1.1 mm, 2.0 mm, 4.2 mm, and 6.2 mm at sufficiently small angles to the (110) planes that their motion effectively is governed by the continuum crystal potential. The
experiment covers a new regime of interaction where each positron emits several
hard photons, whose recoil are not negligible and which are formed on lengths 
where the variation of the crystal field cannot be ignored. As a result neither
the single-photon semiclassical theory of Baier et al. nor the conventional
cascade approach to multiple hard photon emissions (quantum radiation reaction) 
based on the local constant field approximation are able to reproduce the 
experimental results. After developing a theoretical scheme which incorporates
the essential physical features of the experiments, i.e., multiple emissions,
photon recoil and background field variation within the radiation formation
length, we show that it provides results in convincing agreement with the 
data.
\end{abstract}
\maketitle

Strong electromagnetic fields as those produced by intense lasers and by crystals are a unique tool to test QED in the laboratory in unprecedented high-energy regimes, where nonlinear effects in the electromagnetic field energy density dominate the dynamics \cite{Mitter_1975, Ritus_1985,Baier1998,Di_Piazza_2012,Uggerhoj_2005,Fuchs2015}. When electrodynamical processes occur in the presence of a sufficiently intense background electromagnetic field, the photon density of the latter is so high that charged particles like positrons (charge $e$ and mass $m$, respectively) interact coherently with several background field photons. The theoretical description of this regime, known as strong-field QED (SFQED), relies on Lorentz- and gauge-invariant parameters, which depend on the structure of the external electromagnetic field \cite{Landau_b_4_1982}.

When a high-energy positron impinges onto a crystal along a symmetry plane of the crystal lattice, its motion can become transversely bound between two adjacent planes, and the positron experiences an effectively static ``continuum'' potential varying only along the direction perpendicular to the planes (planar channeling), \cite{Lind65,Andersen} and \cite{Baier1998,Uggerhoj_2005,Akhiezer_b_1996}. In planar channeling the transverse motion decouples from the motion along the $y$-$z$ plane and the condition for a positron with initial energy $\varepsilon_0\gg m$ to be channeled within two planes is that the total kinetic plus potential energy $\varepsilon_x=p^2_x(t)/2\varepsilon_0+U(x(t))$ associated to the transverse motion is smaller than the potential energy height $U_0$ between the two planes, \cite{Lind65,Andersen} and \cite{Baier1998,Uggerhoj_2005,Akhiezer_b_1996}. Here, we have assumed that $x$ is the coordinate perpendicular to the symmetry planes and that $p_x\ll \varepsilon_0$ is the positron momentum along that direction (units with $\hbar=c=1$, $\alpha=e^2$ are employed). The condition for planar channeling can be expressed as a bound on the maximal positron angle $\theta$ to the plane while in the crystal, that has to be smaller than $\theta_c\equiv\sqrt{2U_0/\varepsilon_0}$, \cite{Baier1998}. The continuum approximation also applies for $\theta\gtrsim\theta_c$ if $\varepsilon_0 \gg m$.

The study of SFQED processes in the background crystal field corresponding to the continuum potential, such as the emission of high-energy photons is complicated by the necessity of including the field exactly in the calculations. Now, in the case of planar channeling the crystal field has a dependence on the coordinate $x$, which does not allow for an exact analytical solution of the Dirac equation \cite{Wistisen2019}. For this reason the semiclassical method of Baier and Katkov \cite{Baie89a,Baier1998}, which allows for the computation of the probabilities of quantum processes using only the classical trajectory of the charged particles involved in the process, has been extremely useful in the study of SFQED processes. The semiclassical method is based on the observation that in the interaction of ultrarelativistic particles (we consider positrons here) the quantization of the motion is negligible such that one can still attribute physical meaning to the positron classical trajectory, whereas the main quantum effect in the process of radiation to be included is the recoil in the emission of high-energy photons \cite{Baier1998}. The semiclassical method has been successfully employed to compute the probability of the basic SFQED processes like single photon emission and electron-positron photoproduction in aligned crystals (see Ref. \cite{Baier1998} also for studies on higher-order processes). However, when  a positron crosses a crystal whose thickness corresponds to several radiation lengths, a potentially large number of photons can be emitted. The theoretical investigation of such high-order processes is a formidable task \cite{Wistisen2019doublephot}, and mainly kinetic approaches are employed, where it is assumed that multiple photon emissions arise from sequential (cascade) emissions of single photons, each single photon emission being well localized \cite{Baier1998}. The localization of the emission is a crucial requirement of the method and it corresponds to assuming that the formation length $l_f$ of the photon emission process is much smaller than the typical length where the crystal field significantly varies, such that the local value of the probability per unit time, evaluated for a constant field, can be employed \cite{Baier1998,Ritus_1985}. This ``local constant field approximation (LCFA)'' is another remarkable tool in strong-field physics and recent studies have been devoted to investigating its limitations especially in the realm of SFQED in beamstrahlung \cite{Blankenbecler_1996,PhysRevD.92.045045}, in intense laser fields \cite{Di_Piazza_2018c,Blackburn_2018,Di_Piazza_2019d,Ilderton_2019b} and in space-time dependent electric fields \cite{Alexandrov_2019}. The LCFA has previously been applied to high-energy radiation and pair-production processes in aligned single crystals. In Refs. \cite{Baier_1989, Khokonov_2002} the leading-order correction in the field derivatives of the photon radiation probability has been found. In Refs. \cite{BELKACEM1986211,BAK1985491} experimental results beyond the LCFA are presented but either quantum radiation reaction effects were negligible, i.e., each charge emits on average one photon or, in case of multiple photon emission, single-photon spectra were not measured. See Refs \cite{PhysRevLett.111.255502,shul1978suppression,FOMIN1979131} for other channeling related effects.

Here we report experimental single photon spectra emitted by high-energy (50 GeV)
positrons crossing silicon crystals of different thicknesses (1.1 mm, 2.0 mm, 
4.2 mm, and 6.2 mm) aligned to the $(110)$ planes. Depending on the crystal thickness, 
several photons are emitted by each positron with significant recoil, such that 
quantum radiation reaction effects have to be taken into account (see also the 
results of our previous experiment reported in Ref. \citep{Wistisen2018experimental}). 
By employing a conventional kinetic approach based on the emission probabilities
evaluated within the LCFA, we show that such an approach is unable to
explain the experimental results. Thus, we have developed a kinetic approach
particularly suitable for SFQED processes in aligned crystals and where 
effects beyond the LCFA are implemented, see supplementary material \cite{supplmat}. The theoretical spectra
obtained with this method result in overall good agreement with the
data, which in turn can be interpreted as the first experimental
investigation of quantum radiation reaction beyond the LCFA.

\begin{figure}[t]
\includegraphics[width=1\columnwidth]{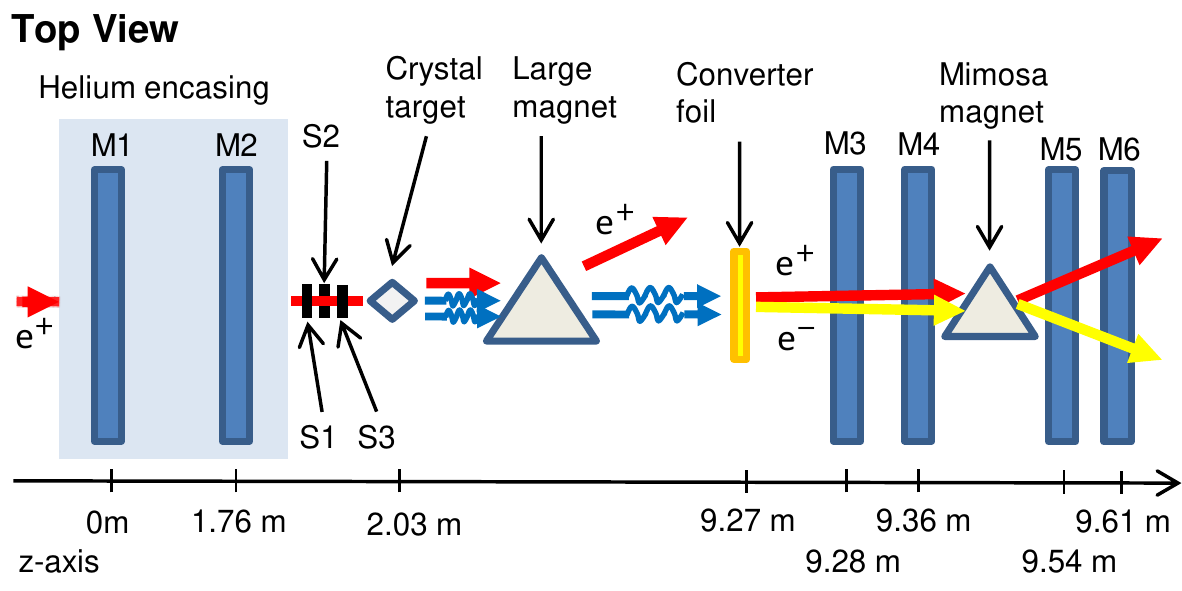}
\caption{A top view schematic of the experimental setup.\label{fig:expfig}}
\end{figure}

The experiment was carried out at the CERN SPS H4 beamline employing 
a positron beam of $50$ GeV with an energy spread of a few percent 
(see Fig. \ref{fig:expfig}).
Due to changed conditions of the accelerator which can occur during 
outages, the positron beam features varying initial angular distributions 
along the $x$-direction (the crystal symmetry planes are defined to 
be parallel to the $y$-$z$ planes). The experimental angular distributions 
were fitted with a Gaussian function and the resulting average 
angle $\theta_0$ relative to the crystal plane and standard deviation $\sigma_{\theta}$ are 
reported in Table \ref{tab:params} for the crystal thicknesses
used in the experiment. As the radiation spectra are highly sensitive to 
the entry angle of the positrons, the variation in the entry angles complicates a 
direct comparison between spectra for the different crystal thicknesses.
\begin{table}
\begin{tabular}{|c|c|c|c|c|c|}
\hline 
$L[\text{mm}]$ & 1.1 & 2.0 & 4.2 & 6.2 & 6.2\tabularnewline
\hline 
\hline 
$\sigma_{\theta}[\mu\text{rad}]$ & 85 & 100 & 85 & 85 & 100\tabularnewline
\hline 
$\theta_0[\mu\text{rad}]$ & 27 & 70 & 62 & 7 & 50\tabularnewline
\hline 
\end{tabular}

\caption{Average angles $\theta_0$ and standard deviations $\sigma_{\theta}$ 
of the Gaussian functions fitting the initial angular distribution of the
positrons along the $x$-direction for the various crystal thicknesses
used in the experiment. For the case of our experiment we have $\theta_c = 30$ $\mu$rad, 
that is, only a minor fraction of positrons is channeled. Note that the critical 
angle $\psi_p$, as given, e.g., in Ref. \cite{Uggerhoj_2005}, assumes the value 
$\psi_p = 23$ $\mu$rad. \label{tab:params}}
\end{table}

The scintillators S1, S2 and S3 are used to make the trigger signal, for
which a signal must be present in S1 and S3 and absent in S2, as S2
has a hole to allow particles through. After the scintillators the positron
enters a Helium chamber to reduce multiple Coulomb scattering. Here,
the transverse position of the positron is measured, as it passes through 
the MIMOSA detectors M1 and M2, which allows to determine the incidence
angle. The positron then enters the silicon crystal, where it emits 
radiation. A large magnet removes the charged particles exiting the crystal.
The emitted photons, instead, continue forward and encounter a 
converter foil of $200\;\text{$\mu$m}$ tantalum, corresponding to an 
approximately $1/26$ chance of being converted into an electron-positron
pair. Note that when the photon energy exceeds the threshold of pair
production, the pair production cross section quickly tends to a 
constant value for large photon energies. This approach of conversion
is used to obtain a spectrum of individual photons, as opposed to
a calorimeter setup, which would only measure the sum of energies of
all the emitted photons. The produced electron and positron pair is
tracked through detectors M3 and M4, are then deflected by another
magnet, and then tracked again in M5 and M6. The deflection angle of
the electron and the positron allows to determine their individual 
momenta, whose sum yields the momentum of the original photon (see 
Ref. \citep{Wistisen2018experimental} for a description of the employed 
tracking algorithm). The response of the experimental setup
is complicated by multiple scattering through the setup, finite detector sizes etc. 
and therefore it should be simulated \citep{Wistisen2018experimental}. 
In order to validate the simulation of the experiment, the crystal can be 
oriented far away from any low-index crystallographic direction, such that 
the emission of radiation essentially stems from Bethe-Heitler (BH) bremsstrahlung 
rather than showing coherence effects as in coherent bremsstrahlung or channeling radiation. 
BH bremsstrahlung is a well studied process, and the agreement between the 
simulation using the BH spectrum and the experimental spectra
shows that the experimental setup is well understood and described
by the simulation. An overall normalization constant is used on the
simulation such that the BH simulation matches the experiment, and
this accounts for the inherent efficiency in the MIMOSA detectors.

\begin{figure}
\includegraphics[width=1\columnwidth]{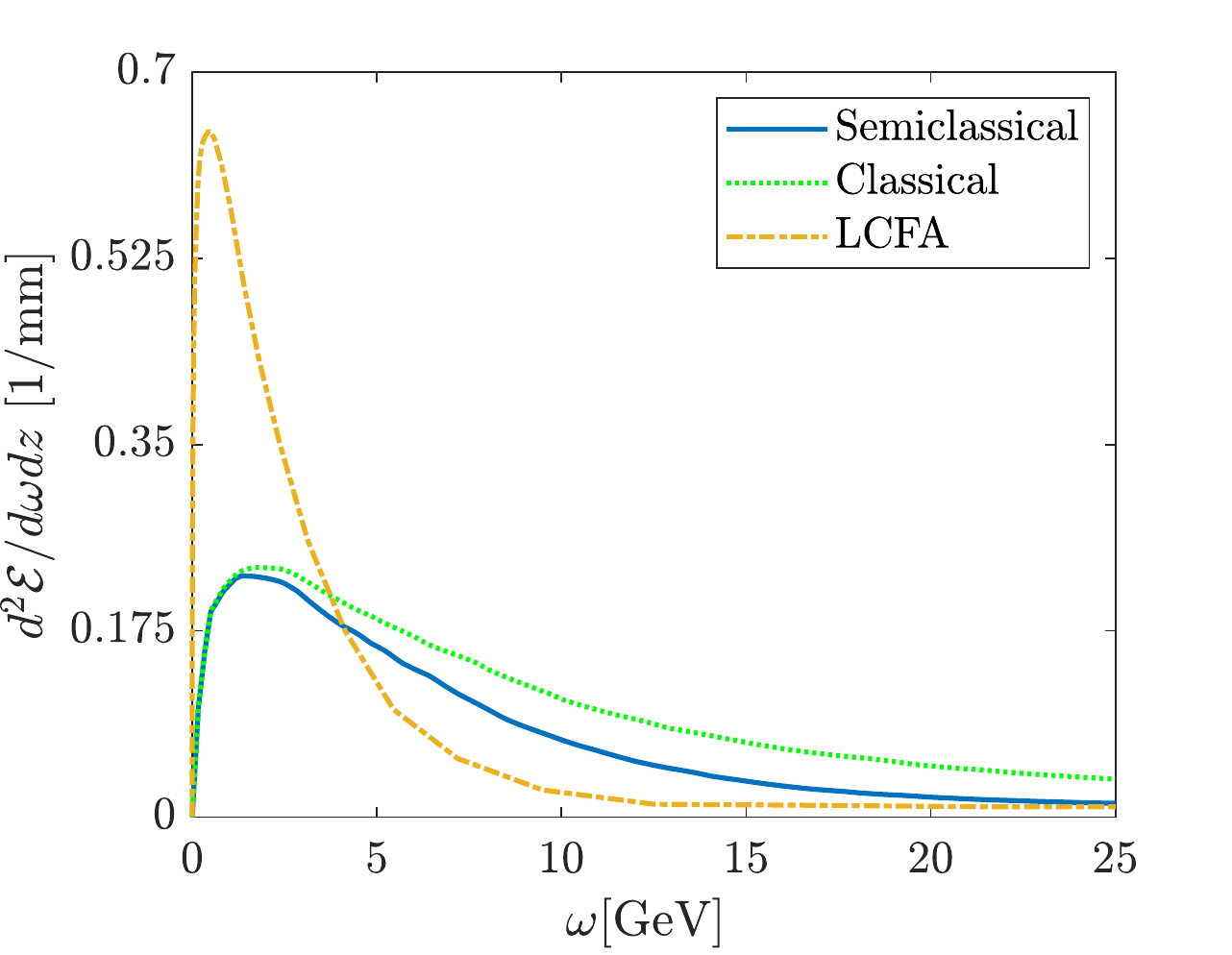}

\caption{Theoretical single-photon power spectra for a positron beam of 50 GeV
energy crossing a crystal of 0.1 mm thickness, which is the energy emitted per 
unit photon energy and per unit length of the crystal. For the incoming
positron angular distribution we have chosen $\theta_0=7\;\text{$\mu$rad}$
and $\sigma_{\theta}=85\;\text{$\mu$rad}$. The green, dotted curve
corresponds to the classical emission spectrum, the blue, continuous curve
correspond to the quantum emission spectrum via the semiclassical
method, and the yellow, dashed curve corresponds to the quantum
emission spectrum within the LCFA. See also Fig. 1 in Ref. \citep{Khokonov_2002}.\label{fig:figthin}}
\end{figure}

For the theoretical description of the experimental results it is useful 
to introduce the parameters $\chi=e\langle\varepsilon E\rangle/m^{3}$
and $\xi=|\bm{p}_{\perp,\text{max}}-\langle \bm{p}_{\perp}\rangle|/m$ \cite{Mitter_1975, Ritus_1985,Baier1998,Di_Piazza_2012,Uggerhoj_2005}. 
Here, $\varepsilon(t)$ is the positron energy at time $t$, $E(t)$ is the amplitude of the crystal electric 
field at the positron position at time $t$, the symbol $\langle\rangle$ indicates the average 
over the positron trajectory, and $|\bm{p}_{\perp,\text{max}}|$ is the maximum momentum transverse 
to the direction of the largest component of the momentum $p_z(t)\approx\varepsilon(t)$
(note that for channeled positrons $\langle p_x\rangle =0$). 
When $\chi$ is of the order of unity or larger, quantum effects such as 
spin and recoil during the emission are essential. The parameter $\xi$ 
differentiates between regimes of undulator-like ($\xi\ll 1$) and 
synchrotron-like ($\xi\gg1$) radiation emission. Quantum radiation 
reaction is the emission of multiple photons while quantum effects in 
the emission is important, i.e. $\chi$ is not too small \cite{PhysRevLett.105.220403}. 
When $\xi\gg1$ the calculation of the quantum radiation reaction process 
is simplified as one can assume that the multiple emissions mainly stem 
from a sequence of localized single-photon events and use the LCFA to 
calculate the corresponding single-photon radiation emission probability. 
This approach has been employed to explain recent experimental results 
on radiation reaction \citep{Wistisen2018experimental,PhysRevX.8.011020,PhysRevX.8.031004}.
When $\xi$ is on the order of unity, the above approach is no longer 
applicable, and a more general theory of radiation reaction is required. 
In the experiment reported here, we have that $\xi<2.9$ and $\chi<0.042$,
with the inequalities being due to different initial conditions
of the positron yielding different values of the parameters. While 
the value of $\chi$ is smaller than unity, it is large enough that 
quantum effects are important in the experiment. This important 
point is illustrated in Fig. \ref{fig:figthin} by a direct 
comparison of classical and quantum spectra of radiation emission 
(average energy radiated per unit of photon energy and unit length)  
for a thin crystal, such that radiation reaction effects, i.e.,
multiple photon emission, can be neglected. 

\begin{figure}
\includegraphics[width=1\columnwidth]{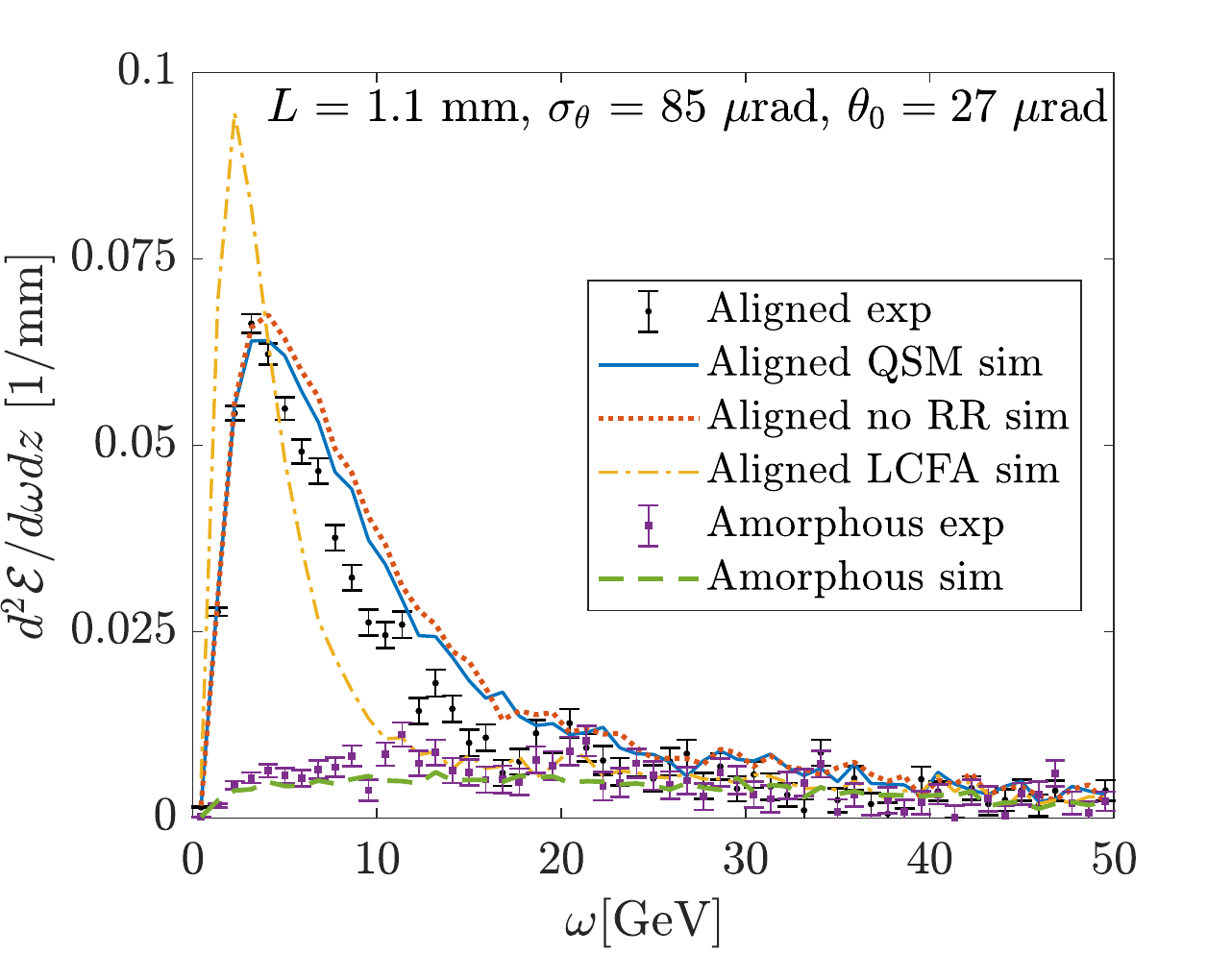}

\caption{Experimental data compared with the three theoretical models described
in the text for the 1.1 mm case.\label{fig:fig1mm}}
\end{figure}

\begin{figure*}
\includegraphics[width=2\columnwidth]{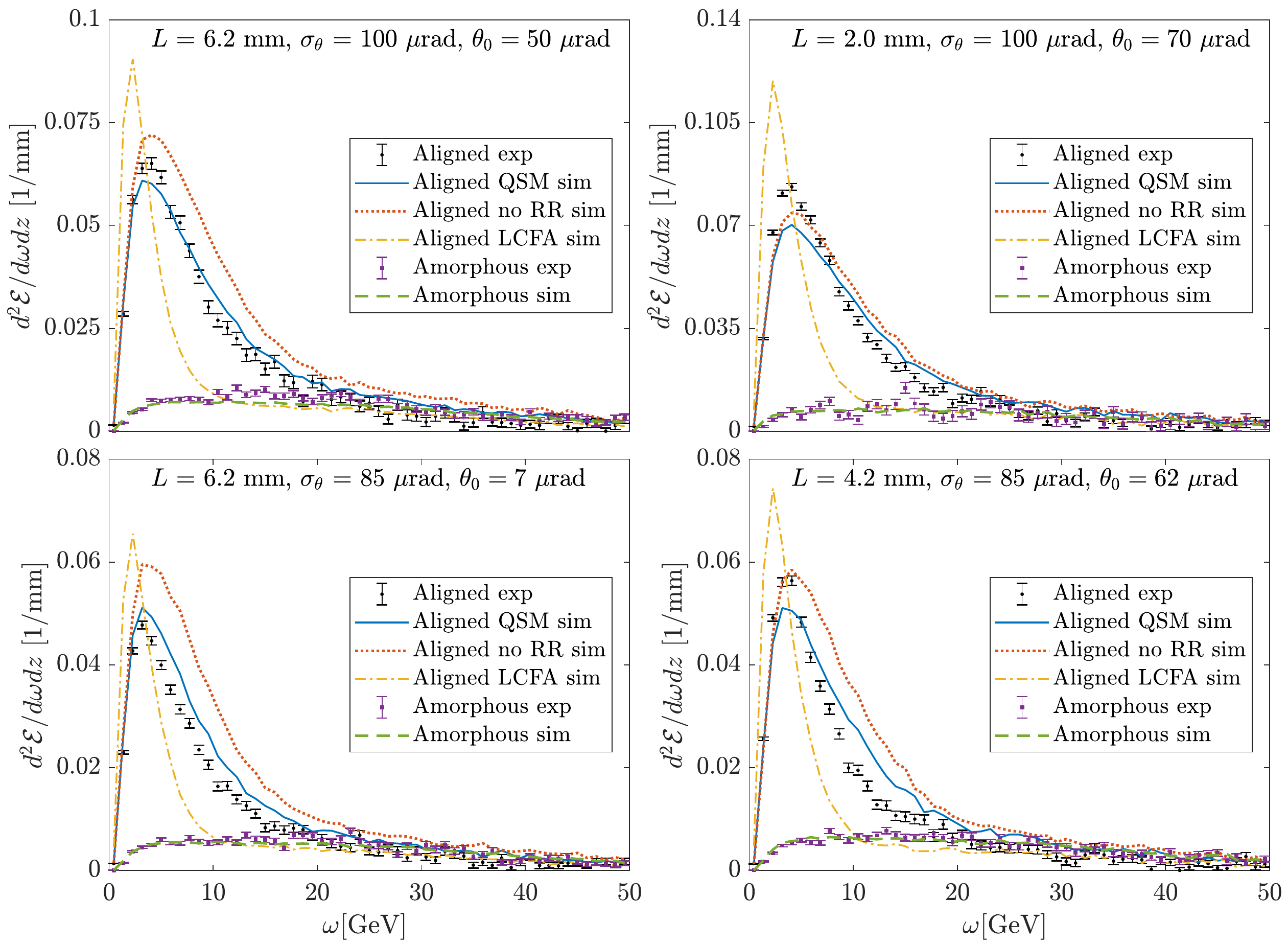}
\caption{Experimental data compared with the three different models 
described in the text, in the `aligned' case. The data with the 
crystal turned into the `amorphous' orientation is compared 
with the simulation of the BH bremsstrahlung spectrum.\label{fig:figall}}
\end{figure*}

In Figs. \ref{fig:fig1mm} and \ref{fig:figall} we show all the 
experimental data corresponding to five different settings
of crystal thickness and beam distributions, see Table \ref{tab:params}.
In all figures we also report the simulation corresponding to
the `amorphous' orientation and we always find a very good agreement 
with the BH bremsstrahlung.

Since quantum effects are important, we compare the experimental data with 
three theoretical quantum models. In the first and most general model,
called quantum stochastic model (QSM), the multiple photon emission is
treated as a cascade of sequential single-photon emissions. Each single-photon
emission event is implemented via a Monte Carlo approach based on positron spin- and
photon polarization-averaged emission probabilities. The new feature of this model 
is the use of the semiclassical method of Baier and Katkov to calculate the 
differential single-photon emission rate $dW/d\omega$ \citep{baier1968processes,Baier1998,Belkacem198586,KIMBALL19861,PhysRevD.90.125008,PhysRevD.92.045045} integrated over a finite section of the positron 
trajectory corresponding to a finite time interval $T$. This approach is then able 
to handle quantum radiation reaction beyond the LCFA for planar channeling and 
takes advantage of the regular, oscillatory motion of positrons inside the crystal 
field. In fact, the value of $T$ has to be large enough such that the differential 
rate $dW/d\omega$ has converged, i.e. it no longer changes significantly when $T$ 
is further increased. This requires a value of $T$ on the order of several photon
formation lengths $l_{f}=2\gamma^{2}(1-\omega/\varepsilon)/\omega$ \citep{Baier1998}, where 
$\gamma=\varepsilon/m$  is the Lorentz factor of the positron at the moment of emission.
We refer to the supplemental material \cite{supplmat} for more details on this scheme.

The second model is the ``LCFA'', which is the usual approach to quantum 
radiation reaction, where multiple photon emissions are simulated
via independent and random emission events, the emission probability 
being used within the LCFA \citep{Wistisen2018experimental}. Finally,
the third model is the ``no RR'', where radiation reaction is
ignored, which is the same as the first model, except that the momentum of
the emitted photon is not subtracted from the radiating positron.
The difference between the first and the third model, therefore, shows
the size of radiation-reaction effects. In Ref. \citep{Wistisen2018experimental}
we described how to use the constant field approximation in the case
of channeling radiation and therefore we refer to this paper for additional
details. The only difference in the LCFA model as compared to that employed
in Ref. \citep{Wistisen2018experimental}, is that here we also add the 
incoherent BH bremsstrahlung with a Monte Carlo approach. The reason is that 
this process is more important here than in Ref. \citep{Wistisen2018experimental}. 

Figures \ref{fig:fig1mm} and \ref{fig:figall} show that the three models give 
three distinctive curves, which means that we are able to distinguish between 
the two models of radiation reaction (within and beyond the LCFA) as well as to 
establish that radiation reaction is present, as otherwise the ``no RR'' curve  
would coincide with the ``QSM'' curve. It is then clearly seen that the LCFA 
is not applicable in the parameter regime under investigation, while the 
QSM model is overall in good agreement with the experimental data.

The process of multiple elastic scattering of the positron with the nuclei
as the positron propagates through the crystal plays an important role
as this on average increases the otherwise conserved energy 
$\varepsilon_x=p_{x}^{2}(t)/2\varepsilon_0+U(x(t))$ associated with the
motion along the $x$-direction. Since
the radiation emission spectrum depends on this effect, we have implemented 
it in our numerical codes as described in \citep{Babaev2012}. The
positron velocity at each timestep, as provided by the solver of the 
trajectory according to the Lorentz force, is additionally changed by 
an amount which is random and normal distributed, with a
standard deviation depending on the local density of nuclei and
electrons. In Ref. \citep{XAVIER1990278} a method similar to what
we have described here was put forward. However, the method was compared
to experimental results with $\xi\gg 1$, where the LCFA was a good approximation. 
Moreover, an important difference between the two methods is that in 
Ref. \citep{XAVIER1990278} the trajectory is divided into sections with length 
of the order of the period of motion which in general does not lead to convergence of
the differential rate (this is, however, acceptable at $\xi\gg1$, for which it was 
applied, because in this case the formation length is generally shorter than the 
oscillation period).


\begin{acknowledgments}
T. Wistisen was supported by the Alexander von Humboldt-Stiftung except for the initial part of the
project where he was supported by the VILLUM FONDEN (research grant VKR023371). 
Except for the first author, the author list is alphabetical and the contribution
of each author was the following: T. N. Wistisen and U. I. Uggerh{\o}j conceived 
and designed the experiment. C. F. Nielsen carried out the data analysis 
with assistance from U. I. Uggerh{\o}j and A. H. S{\o}rensen. C. F. Nielsen, 
U. I. Uggerh{\o}j, A. H. S{\o}rensen and T. N. Wistisen participated in the 
experiment. T. N. Wistisen proposed the theoretical models in collaboration 
with A. Di Piazza and carried out the numerical calculations. 
T. N. Wistisen and A. Di Piazza wrote the paper with input from the rest of the 
collaboration and C. F. Nielsen 
produced the figures with the experimental data.
\end{acknowledgments}

\bibliography{biblio,Books,Papers_Crystal,Papers_PP_and_Cascades,Papers_Radiation,Papers_RR,Papers_Various,Reviews}

\end{document}